\documentclass{PoS}

\def \th {\thinspace}

\def\approxgt{\mathrel{\hbox{\rlap{\lower.55ex \hbox {$\sim$}} \kern-.3em \raise.4ex \hbox{$>$}}}}
\def\lesssim{\mathrel{\hbox{\rlap{\lower.55ex \hbox {$\sim$}} \kern-.3em \raise.4ex \hbox{$<$}}}}
\def\approxlt{\mathrel{\hbox{\rlap{\lower.55ex \hbox {$\sim$}} \kern-.3em \raise.4ex \hbox{$<$}}}}

\title{A model for the Z-track phenomenon, jet formation and the 
kilohertz QPO based on Rossi-XTE observations of the Z-track sources}

\ShortTitle{Jet formation in the Z-track sources}

\author{\speaker{M. J. Church,$^{1,2}$} M. Ba\l uci\'nska-Church,$^{1,2}$ 
       N. K. Jackson$^1$ and A. Gibiec$^2$\\
\llap{$^1$}University of Birmingham,\\
           Birmingham B15 2TT, UK\\
\llap{$^2$}Astronomical Observatory,\\
          Jagiellonian University,\\
          Ul. Orla 171, 30-244, Cracow, Poland\\
Email: \email{mjc@star.sr.bham.ac.uk}, \email {mbc@star.sr.bham.ac.uk},
       \email{njackson@star.sr.bham.ac.uk}, \email{gibiec@oa.uj.edu.pl}}
\abstract{
We present a new model for the Z-track phenomenon, based on analysis of the spectral 
evolution around the Z-track in several Z-track sources, in which radiation
pressure plays a major role. Increasing mass accretion rate on the normal branch
causes heating of the neutron star with the emissive flux from the surface
increasing by an order of magnitude to become super-Eddington at the horizontal
branch where radio detection shows the presence of jets. We propose that the radiation
pressure disrupts the inner disk leading to the launching of the jets. Secondly,
by timing analysis of the same data we find a correlation of the frequency of
kHz QPO with the emissive flux and propose that the higher frequency QPO is an
oscillation at the inner disk edge which progressively moves to larger radial
positions as the disk is disrupted by radiation pressure.}

\FullConference{VII Microquasar Workshop: Microquasars and Beyond\\
		 September 1-5 2008\\
		 Foca, Izmir, Turkey}

\begin{document}

\section{Introduction}

The Z-track sources form the brightest group of Low Mass X-ray Binaries (LMXB) containing a neutron
star, with luminosities at or above the Eddington Limit. They are characterised by having three
distinct branches in a hardness-intensity diagram: the horizontal branch (HB), the normal branch (NB)
and the flaring branch (FB) (Hasinger \& van der Klis 1989) showing that major physical changes take
place within the sources but the nature of these has not been understood. It has been widely thought
that the physical and spectral changes are driven by variation of a single parameter along the Z-track,
presumably the mass accretion rate (Priedhorsky et al. 1986), assumed to increase monotonically
in the direction HB - NB - FB. However, the evidence for this is rather limited based mostly
on the assumed identification of increased UV emission in a multi-wavelength campaign on Cyg\th X-2
with the flaring branch (Vrtilek et al. 1990) which has since been questioned (Church et al. 2006).
Moreover the variation of the X-ray intensity does not obviously support this as the intensity
does not increase monotonically in the direction HB - NB - FB, but decreases
on the normal branch. Arguably the most important feature of the Z-track sources is the detection
of radio emission showing jets to be present but essentially only in the upper normal and horizontal branches
(e.g. Berendsen et al. 2000). The presence of jets was dramatically demonstrated by extended radio observations
of the Z-track source \hbox{Sco\th X-1} (Fomalont et al. 2001) which revealed radio condensations moving
away from the source with velocity $v/c$ of 0.45. Thus the Z-track sources uniquely provide the
possibility of determining the physical conditions within the sources on the part of the
Z-track where radio is detected so telling us the conditions needed for the launching of jets.
Apart from this, an understanding of the Z-track sources is essential to the basic understanding of
LMXB in general. Extensive work has been carried out on the timing properties of Z-track sources (van
der Klis et al. 1987; Hasinger \& van der Klis, 1989) and revealed the existence of quasi-periodic
oscillations (QPO) which change along the Z-track. However, analysis of the timing properties has not
provided an explanation of the Z-track phenomenon. Spectral analysis is more likely to reveal the
nature of the physical changes taking place as directly showing changes in the emission components
during the spectral evolution along the Z-track, but spectral studies of the sources have been
hindered by lack of agreement over the emission model to be used.

\begin{figure*}[!hb]                                                      
\begin{center}
\includegraphics[width=140mm,height=45mm,angle=0]{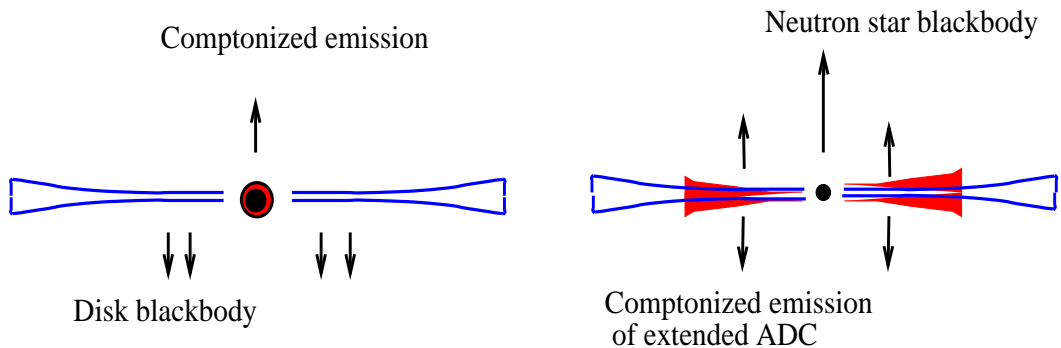}                   
\caption{Left: the Eastern model for LMXB; right: the Extended ADC model}
\end{center}
\end{figure*}

The spectra of LMXB clearly consisting of a power law Comptonized component plus a thermal blackbody 
can be interpreted using two very different physical models (Fig. 1). In the ``Eastern'' model
(Mitsuda et al. 1989), the thermal emission is multi-colour disk blackbody and the non-thermal
\begin{figure*}[!hb]                                                      
\begin{center}
\includegraphics[width=70mm,height=100mm,angle=270]{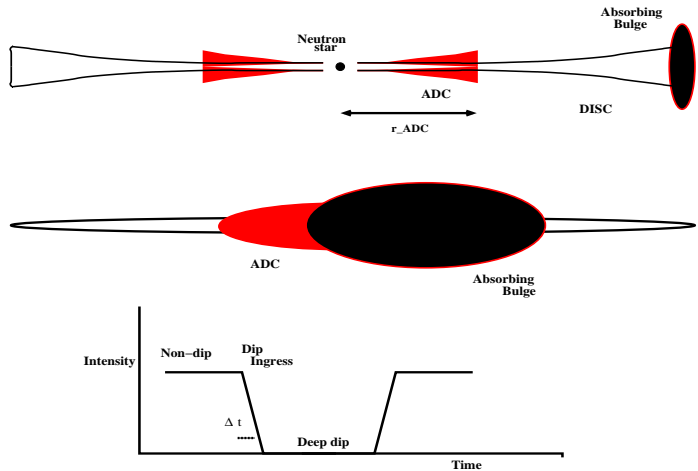}
\caption{The technique of dip ingress timing: an extended ADC is viewed  in elevation (top)
and along the line-of-sight to the observer (middle), showing that the dip ingress time
depends on the radial size of the extended region, so that measurement of the ingress time
from the lightcurve (bottom) provides the radial extent of the ADC.}
\end{center}
\end{figure*}
component is Comptonized emission from a small central region: the neutron star atmosphere
or the inner disk. In the ``Extended ADC'' model (Church \& Ba\l uci\'nska-Church 2004),
the neutron star is the source of blackbody emission and an extended accretion disk corona
produces the Comptonized 
emission. The dipping class of LMXB provide strong evidence for the very extended nature of
the ADC since dip ingress timing as illustrated in Fig. 2 allows determination of the radial extent 
of the ADC 
$r_{\rm ADC}$, using the dip ingress time $\Delta$t, the orbital period $P$ and the accretion 
disk radius $r_{\rm AD}$ via
\[{2\, r_{\rm ADC}\over \Delta t} = {2\, \pi \, r_{\rm AD}\over P}.\]

By applying this technique to most of the dipping LMXB, it was found that the ADC was indeed
very large, typically 50\th 000 km in radial extent, i.e. $\sim$ 15\% of the accretion disk size, 
and varying linearly with source luminosity
between 20\th 000 and 700\th 000 km (Church \& Ba\l uci\'nska-Church 2004). This shows that
the accretion disk corona is as shown schematically in Fig. 2, i.e. an extended, thin ADC 
(having $H/r$ $<$ 1) above the accretion disk.

Strong independent 
support comes from {\it Chandra} grating results for Cygnus\th X-2 (Schulz et al. 2008) in which 
broad emission lines of highly ionized species were located at radial distances of 20\th 000 
- 110\th 000 km in an extended ADC. Theses results rule out the Comptonizing region being a
small central region, and so the Eastern model.

The majority of spectral fitting of the Z-track sources has employed the Eastern model: e.g.
Done et al. (2002), Agrawal \& Sreekumar (2003), di Salvo et al. (2002). However, it proved
difficult to interpret the results, and no concensus view of the nature of the Z-track emerged.
We have applied the Extended ADC model to several of the Z-track sources as described below, 
and have found that the spectral fitting results clearly suggest a model for the Z-track
phenomenon, for jet launching and for the nature of the kHz QPO.

\begin{figure*}[!ht]
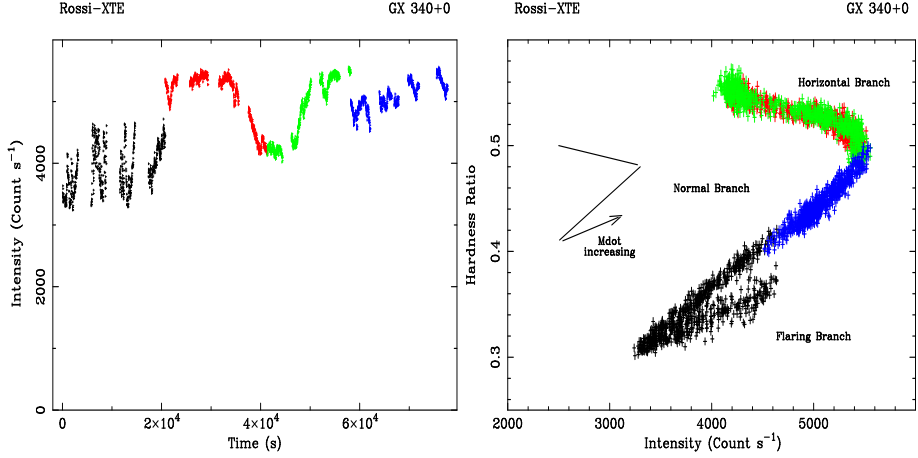
                                                   
\begin{center}
\includegraphics[width=60mm,height=60mm,angle=270]{full_lc}
\includegraphics[width=60mm,height=60mm,angle=270]{full_ztrack}
\caption{{\it RXTE} observations of GX\th 340+0: left: the PCA lightcurve and right: the corresponding Z-track.
The arrow shows the direction of $\dot M$ increase in the standard view of the Z-track sources.}
\end{center}
\end{figure*}

\begin{figure*}[!h]
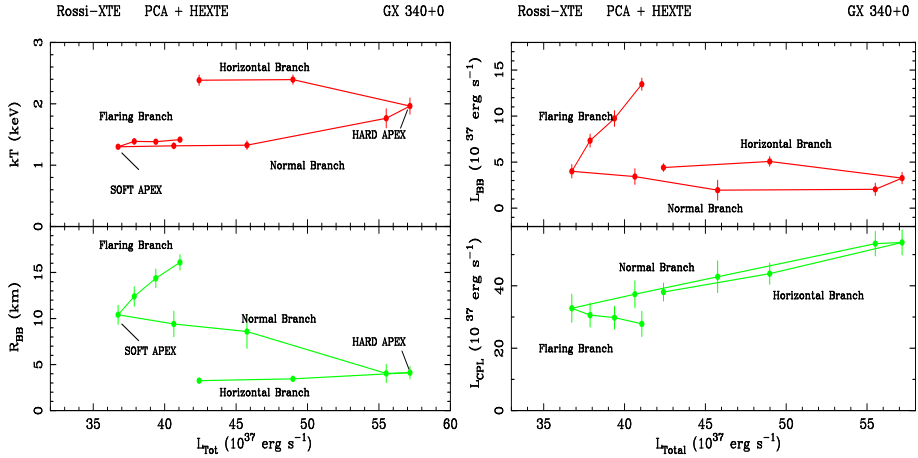
                                                    
\begin{center}
\includegraphics[width=60mm,height=60mm,angle=270]{bb_sub}
\includegraphics[width=60mm,height=60mm,angle=270]{lum_sub}
\caption{Spectral fitting results; left: neutron star blackbody temperature and radius; right:
luminosity of the blackbody (upper) and of the Comptonized ADC emission (lower)}
\end{center}
\end{figure*}

We firstly present results of analysis of high quality {\it RXTE} observation of the Z-track
source GX\th 340+0 (Church et al. 2006). Spectral fitting was carried out using a model
{\sc abs}$\ast$ ({\sc bb + cut}), i.e. a blackbody and a cut-off power law to represent
Comptonization in an extended ADC, for which the seed photons are the disk blackbody emission
(see Church \& Ba\l uci\'nska-Church 2004),
with absorption, and this was done at a sequence of positions around the Z-track. Results for 
the blackbody temperature $kT_{\rm BB}$ and blackbody radius $R_{\rm BB}$ are shown in Fig. 4 (left). 
It can be seen that at the soft apex between NB and FB, $kT_{\rm BB}$ is smallest while $R_{\rm BB}$
$\sim$10.5 km indicates that the whole star is emitting. We propose that at this position, the mass
accretion rate is minimum. Ascending the NB towards the hard apex, $kT_{\rm BB}$ increases
while the blackbody radius falls.  

Fig. 4 (right) shows the 1- 30 keV luminosities of the blackbody $L_{\rm BB}$ and Comptonized ADC emission
($L_{\rm ADC}$). There is a large increase of $L_{\rm ADC}$ moving from the soft apex to the
hard apex, corresponding to the large increase in X-ray intensity. We suggest that this is
strong evidence that $\dot M$ is increasing {\it contrary} to the standard view that $\dot M$
increases around the Z-track in the direction HB - NB - FB.

\section{Spectral analysis of GX\th 340+0, GX\th 5-1 and Cygnus\th X-2}
\subsection{High radiation pressure and jet formation}

\begin{figure*}[!ht]
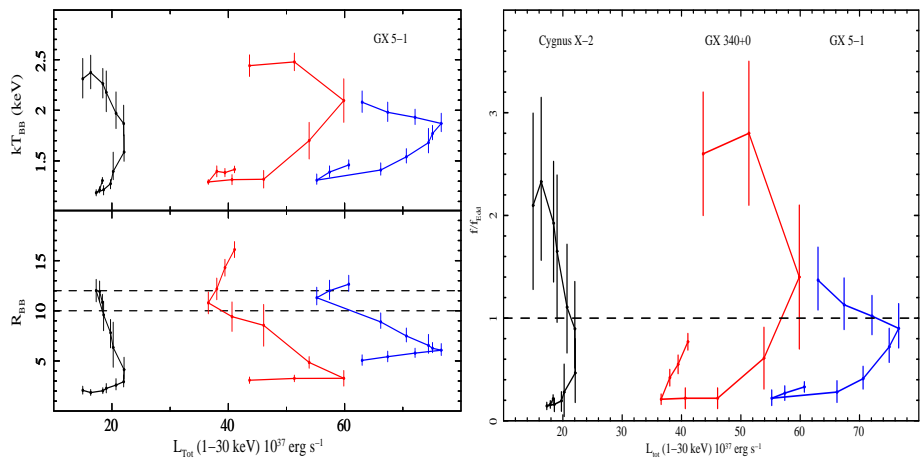
                                                           
\begin{center}
\includegraphics[width=60mm,height=60mm,angle=270]{bb}
\includegraphics[width=60mm,height=60mm,angle=270]{Edd}
\caption{Results for all of the Cygnus\th X-2 like sources; left: the blackbody temperature $kT_{\rm BB}$ 
and radius $R_{\rm BB}$ for the neutron star emission; right: corresponding values of the 
surface emissive flux $f$ expressed in terms of the Eddington value of the flux $f_{\rm Edd}$
(see text).} 
\end{center}
\end{figure*}

We find that all of the three Cygnus\th X-2 like sources: Cygnus\th X-2, GX\th 340+0 and GX\th 5-1 
behave in the same way as seen in Fig. 5 (left), with the neutron star blackbody temperature increasing 
on the normal branch. Similarly all three sources appear 
to emit over the whole surface of the neutron star at the soft apex, and based on that assumption, we 
have a measurement of the radius of the neutron star with a mean of of 11.4$\pm$0.6 km at 90\% confidence 
for the three sources. The increase of $kT$ and decrease of emitting area on the neutron star seen
in moving away from the soft apex towards the hard apex means that the radiation pressure
of the neutron star emission increases substantially. 

Fig. 5 (right) shows the variation
around the Z-track of the ratio of the flux emitted per unit surface area of the neutron star 
to the Eddington flux $L_{\rm Edd}/4\, \pi R^2$, where $R$ is the radius of the neutron star.
This shows in all cases that the sources are emitting at only 20\% of the Eddington flux at the
soft apex, that the emitted flux increases to the Eddington value at the hard apex, and becomes strongly 
super-Eddington on the horizontal branch.

We have thus proposed (Church et al. 2006, 2008) that the strong radiation pressure causes disruption of
the inner accretion disk. The inner disk in these sources has a height $H$ of the order 
of 50 km because the disk is inflated by its own radiation pressure. In this geometry,
the increasing radiation pressure of the neutron star will exert a force horizontally
in to the disk, but will also have a component close to vertical, i.e. acting on material
up to 50 km above the disk. Hydrostatic equilibrium will be destroyed and the accretion 
flow diverted vertically upwards so launching the jet. The detection of radio emission
from the jet correlates very well with the positions on the Z-track where the neutron star
emission becomes super-Eddington. It has, of course, previously been suggested that radiation
pressure be important in jet formation (Begelman \& Rees 1984), and Lynden-Bell (1978)
proposed that a radiatively-supported inner disk would define conical funnels above
and below the disk within which jets may be formed.

The decrease in blackbody radius $R_{\rm BB}$ would be expected from previous results.
Our results in an {\it ASCA}
survey of LMXB (Church \& Ba\l uci\'nska-Church 2001) showed that the neutron star emitting area
in all types of source covering a wide range of luminosities depended on the luminosity of the
source, i.e. the mass accretion rate $\dot M$. This could be viewed geometrically in that the
height of the emitter on the neutron star $h$  was found to agree well with the height of the
inner disk $H$. It was later shown that this behaviour would be expected from the theory of
Inogamov \& Sunyaev (1999) in which the accretion flow adapts to the rotation of the neutron star
in a boundary layer on the star (Church et al. 2002). In the present situation  where the accretion
flow is partly diverted away from the neutron star and the inner disk height very much reduced, the
contraction of the emitting region from the full neutron star to an emitting band at the equator
of height $h$ $<$ 10 km would thus be expected.

\subsection{The nature of the flaring branch}

\vskip 15 mm
\begin{figure*}[!h]                                                     
\begin{flushleft}
\hskip 8mm
\includegraphics[width=60mm,height=35mm,angle=0]{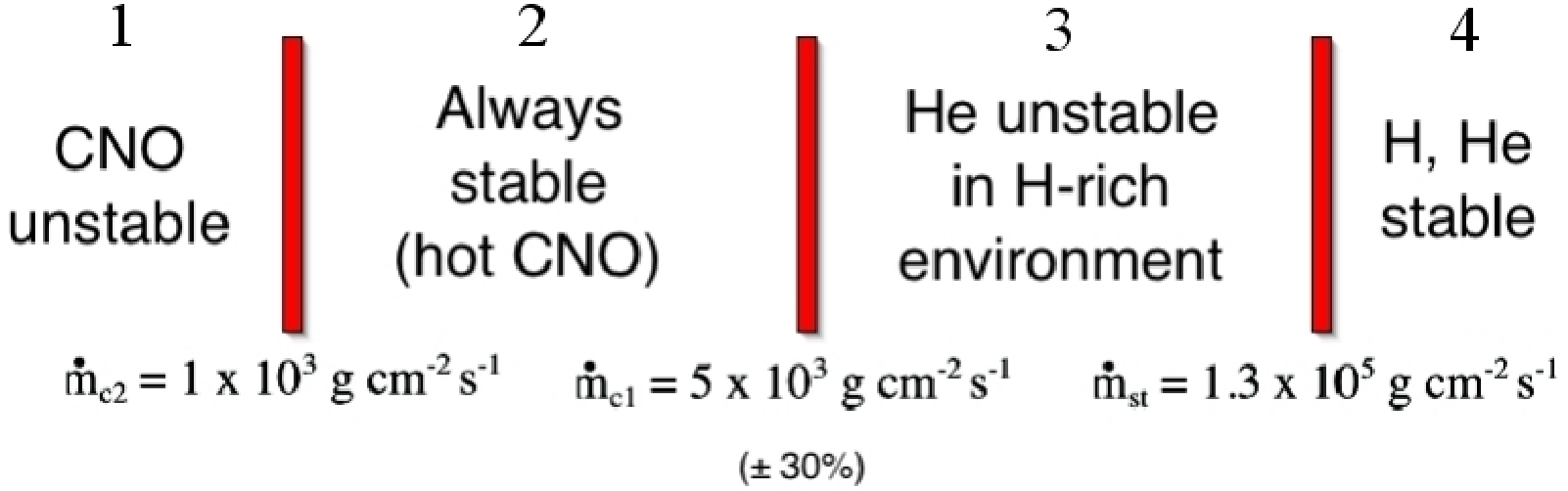}
\caption{Left: the r\'egimes of nuclear burning on the surface of the neutron star demarcated by 
critical values of the mass accretion rate per unit area $\dot m$; right: comparison of measured 
$\dot m$ with the critical value of $\dot m$ (dashed line) that separates the r\'egimes 3 and 4}
\end{flushleft}
\end{figure*}

\vskip -90mm
\begin{figure*}[!h]                                                     
\begin{flushright}                                                      
\hskip -8 mm
\includegraphics[width=60mm,height=60mm,angle=270]{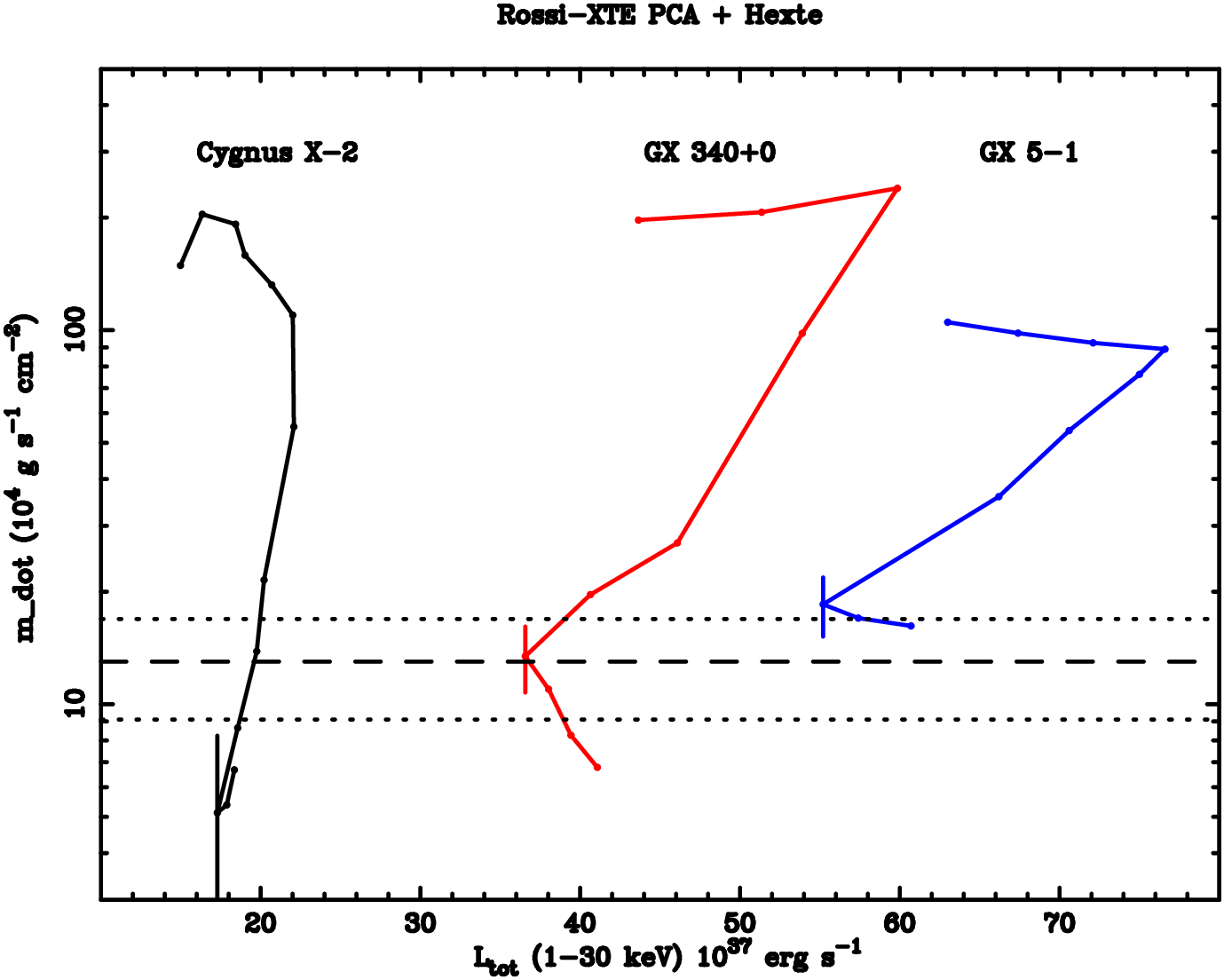}
\end{flushright}
\end{figure*}

\vskip 10mm
The spectral fitting results also provide an explanation of the flaring branch in the Z-track
sources. From Fig. 4 (right) it can be seen that in flaring, $L_{\rm ADC}$ does not increase
suggesting that there is no change in $\dot M$; the blackbody luminosity, however, increases
leading to the overall increase of luminosity in flaring. In Fig. 6 we compare the onset of 
flaring with the theory of unstable burning of the accumulated accretion flow on the surface 
of the neutron star (Fushiki \& Lamb 1987; Bildsten 1998; Schatz et al. 1999). There are
four r\'egimes of nuclear burning demarcated by the value of $\dot m$: the mass accretion rate 
per unit area of the neutron star, and in particular, there is a critical value of $\dot m$ 
which divides r\'egimes 3 and 4 between unstable burning of He in a mixed H/He atmosphere 
and stable burning (Fig. 6 left).
In Fig. 6 (right) we show the measured $\dot m$ obtained from the mass accretion rate $\dot M$ 
divided by the emitting area $4\, \pi \,R_{\rm BB}^2$, also showing the critical value
as the dashed horizontal line with 30\% uncertainties from the
theory. There is good agreement of $\dot m$ at the soft apex suggesting
that as the sources approach this apex along the NB, the surface burning becomes unstable
and $L_{\rm BB}$ increases by the energy release, which is seen as flaring in the lightcurve.
  
\begin{figure*}[!h]                                                   
\begin{center}
\includegraphics[width=60mm,height=60mm,angle=270]{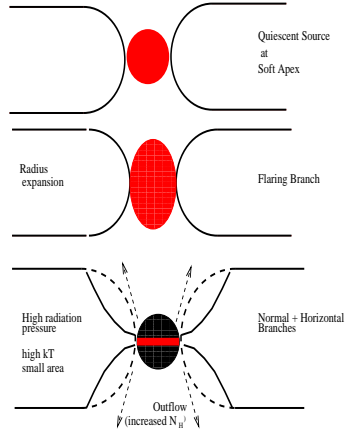}
\caption{Schematic view of the model of the Z-track phenomenon; top: the soft apex;
middle: the flaring branch; bottom: normal and horizontal branches in which the inner disk
is disrupted and a jet forms. In flaring, we show an increase of $R_{\rm BB}$ as measured 
by several authors}
\end{center}
\end{figure*}

We thus propose the model of the Z-track sources and of jet formation summarised in Fig. 7
The quiescent source is shown at the top, with a radiatively-supported thick accretion
disk around a neutron star emitting from its full surface, and with minimum $\dot M$.
In flaring, $\dot M$ is constant, but the neutron star emission increases by unstable
burning, and the blackbody radius may increase to 15 - 18 km suggesting burning expands
beyond the surface of the neutron star as seen in radius-expansion bursts. Such an increase 
was also seen in GX\th 5-1 by Christian \& Swank (1997). Finally,
on the NB and HB, $\dot M$ increases leading to increased blackbody temperature and
radiation pressure that removes the inner disk and diverts the accretion flow into
the vertical direction.

\section{A model for the upper frequency kHz QPO}

\begin{figure*}[!h]
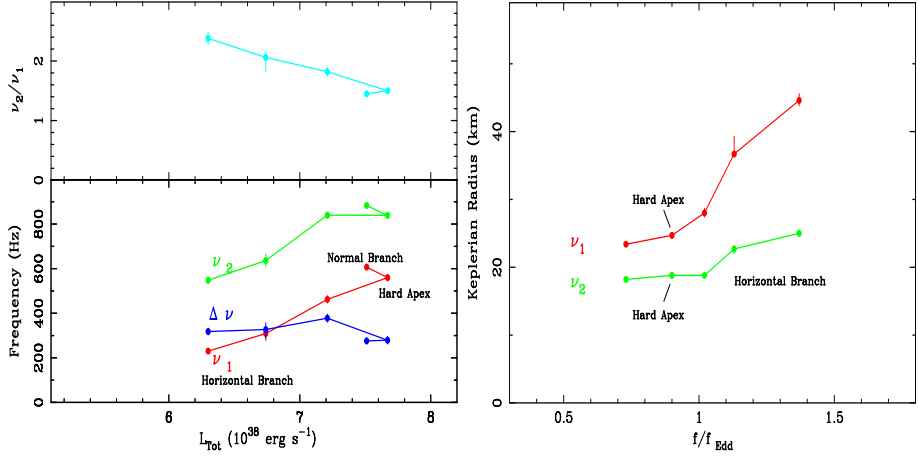
                                                  
\begin{center}
\includegraphics[width=60mm,height=60mm,angle=270]{diff}
\includegraphics[width=60mm,height=60mm,angle=270]{rad_ratio}
\caption{Left: variation of kHz QPO frequencies along the horizontal branch with the 
corresponding variations of the difference $\nu_2$ - $\nu_1$ and of the ratio $\nu_2/\nu_1$;
right: the Keplerian radii of each QPO as a function of the parameter showing the
strength of the radiation pressure $f/f_{\rm Edd}$ (see text)}
\end{center}
\end{figure*}

The above model for the Z-track phenomenon has also led to an interesting result relevant
to the kHz QPO, the nature of which remain unexplained, although it is widely realized
that the upper frequency QPO is likely to relate to orbital motion some way into the
inner accretion disk. We have carried out timing analysis for the same selections of
data as used in the above spectral analysis and now present results in the case of GX\th 5-1.
We used the technique of converting the power spectra into {\it Xspec} format so as to be
able to use the associated powerful spectral fitting software. Fig. 8 (left) shows the results
in a conventional form, firstly as the QPO frequencies $\nu_1$ and $\nu_2$ as a function of 
1- 30 keV luminosity and also as the difference $\nu_2$ - $\nu_1$ and the ratio
$\nu_2/\nu_1$. The kHz QPOs are seen essentially only on the horizontal branch as is
well known.

We also show the results in an unconventional way by plotting the Keplerian radius of each
QPO as a function of $f/f_{\rm Edd}$, the
parameter we have previously used as a measure of radiation pressure in the proximity of the
neutron star (Fig. 8 right), noting that this quantity is {\it measured} from spectral
fitting results, not deduced or interpreted. During movement along the Z-track from the end
towards the hard apex, the upper frequency QPO increases in frequency and so its Keplerian 
radius decreases, from 25 to 18 km. The striking result is that the variation in frequency
as a function of $f/f_{\rm Edd}$ is initially small when the source is sub-Eddington, but
changes rapidly when $f$ is greater than $f_{\rm Edd}$.

There is a clear implication that the oscillation comprising the upper frequency QPO at
$\nu_2$ is {\it always} at the inner edge of the disk, and that the radius of this edge increases
with increasing radiation pressure. Thus the change in QPO frequency is fundamentally driven
by change of the mass accretion rate $\dot M$ in the same way that it is thought that change of
$\dot M$ causes the major physical changes between the Z-track branches, although it is not agreed
in which direction $\dot M$ increases. Sideways movement of the whole Z-track between observations 
is known, forming the so-called ``parallel tracks'', which are not well-understood. However, the 
suggestion that spectral and timing parameters may not therefore be a simple function of
$\dot M$ is neither accepted nor proven and thus the existence of parallel tracks clearly 
does not mean that the changes in QPO frequency are {\it not} caused by changing $\dot M$.

The lower frequency QPO corresponds to Keplerian motion at radial positions
of 22 - 45 km has a variation that follows the form of the higher frequency QPO. The implication
from Fig. 8 (right) 
is that this frequency $\nu_1$ clearly depends on the value of $\nu_2$ although the mechanism is
not clear. The resonance model (Abramowicz \& Klu\'zniak 2001; Klu\'zniak et al. 2007)
or the relativistic model of Stella \& Vietri (1998) are possible; however, the present data do 
not support the sonic point model (Miller \& Lamb 1998) because this model requires formation
of a sonic point in the disk by radiation drag forces, whereas we propose that the disk is
disrupted by radially-outwards radiation pressure, not at a sonic point. Moreover, our model 
does not {\it need} the existence of a sonic point, an Alfv\'en radius or an innermost stable orbit 
to define the inner radius of the accretion disk.

\vfill\eject
\section{Conclusions}

Based on extensive analysis of high quality data of the Cygnus\th X-2 like Z-track sources
we have proposed a model for the Z-track phenomenon which explains the launching of jets
at the particular part of the Z-track on which the radio jets are strongest.
In this model, $\dot M$ increases from the soft apex to the hard apex, i.e. in the opposite
direction to that assumed by many workers as the standard model, although it should be noted
that the evidence for $\dot M$ changing in that direction was never strong. Substantial support
for the model comes from the detected increase in column density in spectral fitting on the NB moving towards
the hard apex in all three sources (Church et al. 2006, 2008), since this increase has to be intrinsic and supports
the disruption of the inner disk. On the flaring branch we show that the results agree very
well with the theory of unstable nuclear burning. Finally, we propose that the upper frequency
kHz QPO is an oscillation that always takes place at the inner disk edge, the radial position
of this varying along the Z-track as the inner disk is disrupted by radiation pressure.
Thus as shown on Fig. 7 (lower) the disrupted disk is essentially cusp-like, and the oscillation
takes place at this cusp which forms a natural site for such an oscillation.

\acknowledgments{This work was supported by the Polish Ministry of
Higher Education and Science grant no. 3946/B/H03/2008/34 and by PPARC grant PPA/G/S/2001/00052.}

\end{document}